\documentclass[a4paper,11pt]{article}
\usepackage{jcappub} 
\usepackage[dvipsnames]{xcolor}
\usepackage[normalem]{ulem}
\usepackage{subcaption}
\usepackage{hyperref}
\usepackage{ulem}
\usepackage{makecell}

\title{Energy spectra of elemental groups of cosmic rays
with~the KASCADE experiment data and machine learning}

\author[a]{M.\,Yu.~Kuznetsov}
\author[a, b, c]{N.\,A.~Petrov}
\author[a, d, e]{I.\,A.~Plokhikh}
\author[a]{V.\,V.~Sotnikov}

\affiliation[a]{Institute for Nuclear Research of the Russian Academy of Sciences, \\
60th October Anniversary Prospect 7a, Moscow 117312, Russia}

\affiliation[b]{Budker Institute of Nuclear Physics SB RAS, \\
Akademika Lavrentieva prospect 11, Novosibirsk 630090, Russia}
    
\affiliation[c]{Department of Physics, Novosibirsk State University, \\
Pirogova street 1, Novosibirsk 630090, Russia}
  
\affiliation[d]{Kutateladze Institute of Thermophysics SB RAS, \\
Akademika Lavrentieva prospect 1, Novosibirsk 630090, Russia}

\affiliation[e]{Department of Mathematics and Mechanics, Novosibirsk State University, \\
Pirogova street 1, Novosibirsk 630090, Russia}

\emailAdd{mkuzn@inr.ac.ru}

\abstract{
We report the reconstruction of the mass component spectra of cosmic rays (protons, helium, carbon, silicon and iron) and their mean mass composition, at energies from 1.4 to 100 PeV. The results are derived from the archival data of the extensive air shower experiment KASCADE. 
We use a novel machine learning technique
developed specifically for this reconstruction, and post-LHC hadronic interaction models: \mbox{QGSJet-II.04}, \mbox{EPOS-LHC} and \mbox{Sibyll~2.3c}. 
We have found an excess of the proton component and a deficit of intermediate and heavy nuclei components compared to the original KASCADE results.
The spectra of protons and helium show a knee-like behavior at $\sim 4.4$~PeV and $\sim 11$~PeV, with significances $5.2\sigma$ and $3.9\sigma$, respectively. 
The spectrum of the iron component has a hint~($2.4\sigma$) of a hardening at $\sim 4.5$~PeV, which can be interpreted as a counterpart of a hardening in the proton spectrum at $166$~TeV, recently reported by the \mbox{GRAPES-3} experiment.
The systematic uncertainties
of our analysis were found to be smaller than those of the original KASCADE, as well as those of IceTop and TALE experiments, over the most part of the energy range studied. We also estimated separately the uncertainty related to the difference between the three mentioned hadronic interaction models.
We also compute a mean logarithm mass of cosmic ray flux as a function of energy. It is in agreement with the results of IceTop, TALE and LHAASO within the uncertainties.
}

\keywords{cosmic ray experiments}

\begin{document}
\maketitle
\flushbottom

\section{Introduction}
\label{sec:intro}

Measurements of the cosmic-ray energy spectrum and the individual mass component spectra in the 1 to 100~PeV energy range provide important information for understanding the sources, acceleration, and propagation mechanisms of galactic cosmic rays~(CRs). They are also essential in searching for the energy region where the transition from galactic to extragalactic CRs takes place.
In particular, many mechanisms of the CR acceleration naturally predict that the maximum energy achievable by cosmic rays is proportional to their charge. This should lead to a sequence of steepenings (``knees'') in the observed spectra of CR mass components, that are coinciding with the maxima of their injected spectra in galactic sources~\cite{1961NCim...22..800P, Gaisser:2011klf, Gaisser:2013bla}. Alternatively, such a behavior of the observed individual spectra can occur due to energy dependent diffusive escape of the CRs from the Galaxy~\cite{Giacinti:2015hva}. Importantly, in both of these models, the features of the observed spectra depend on magnetic rigidity of the respective particles, i.\,e. the energy of the knee in the spectrum of each component is proportional to the charge of this component.

A first indication that such a behavior of CR component spectra is indeed taking place was found by the KASCADE experiment~\cite{kascade_cuts}. Although the increase of the mean mass of the CR flux with energy was supported by other experiments in general, there is a controversy about the particular proportion of individual mass components in the flux depending on energy.
The study of CRs with energies above $\sim 1$~PeV is complicated, as one can only observe it indirectly through extensive air showers (EAS) of secondary particles that they initiate in the Earth's atmosphere. Therefore, the results of various experiments are subject to a number of systematic uncertainties accompanying the reconstruction of primary particle features from the EAS observables.
In addition to the original KASCADE mass component results~\cite{kascade_cuts, Apel:2008cd, Apel:2013uni} we can mention studies of 
other experiments: CASA-BLANCA~\cite{Fowler:2000si},
Tibet AS$\gamma$~\cite{TibetASGamma:2005hrd},
KASCADE-Grande~\cite{Apel:2013uni, Kang:2023lre}, Tunka-133~\cite{Prosin:2014dxa}, TAIGA~\cite{Astapov:2022keo}, IceTop~\cite{IceCube:2019hmk}, and TALE~\cite{TelescopeArray:2020bfv},
in the energy range of interest~\footnote{There is also a study of LHAASO experiment~\cite{LHAASO:2024knt}, that appeared when the present paper was already in the review process.}. There is also another study based on the public KASCADE data~\cite{Arsene:2023cjb}.

In this research, we present mass component spectra reconstruction with machine learning methods using archival data from the KASCADE experiment~\cite{KASCADE:2003swk}, provided by the \mbox{KASCADE} Cosmic Ray Data Centre (KCDC)~\cite{Haungs:2018xpw}.
In the original KASCADE analyses~\cite{kascade_cuts, Apel:2008cd, Apel:2013uni, Finger:2011bia}, the method of two-dimensional unfolding was used, in which a distribution of events over the reconstructed numbers of electrons and muons was converted into a distribution over the type and energy of the primary particle.
In this research, we use a different approach to reanalyze the original data of the KASCADE experiment and to re-derive the CR mass component spectra from it. 
This approach combines the event-by-event classification of the primary particle type with machine learning~\cite{Kostunin:2021eyp, Kuznetsov:2023boz, kgml_method} and two separate unfolding procedures for the correction of the reconstructed primary particle type and energy~\cite{kgml_method}. A detailed description of our methods and their uncertainties was given in Ref.~\cite{kgml_method}.
The additional feature of the present analysis is
the estimation of an independent systematic uncertainty~(called theoretical uncertainty from now on) associated with the difference between
three modern hadronic interaction models (\mbox{QGSJet-II.04}~\cite{Ostapchenko:2010vb}, \mbox{EPOS-LHC}~\cite{Pierog:2013ria} and \mbox{Sibyll~2.3c}~\cite{Riehn:2015aqb}).
All the Monte Carlo simulations used in this study are also provided by the KCDC.

The paper is organized as follows: in Section~\ref{sec:expMC}, we briefly describe the data and Monte Carlo~(MC) that we are using.
Section~\ref{sec:method} outlines the main details of our analysis and also describes some features that were not presented in our methodological paper~\cite{kgml_method}.
Namely, we discuss the estimation of the theoretical uncertainty related to hadronic interaction models and the method of the mean logarithmic mass reconstruction.
In Section~\ref{sec:results}, we show the main results of this study: the reconstructed spectra of individual mass components and the mean logarithmic mass. We also compare these results with other experiments.
The discussion and conclusion are presented in Section~\ref{sec:conclusion}.

\section{Experiment, data and Monte Carlo}
\label{sec:expMC}

In this study, we analyze the archival data from the KASCADE experiment using the machine learning methods developed specifically for this task. The KASCADE air shower experiment was operated from 1996 to 2013 at the KIT Campus in Karlsruhe, Germany (${49.1}^\circ$ north, ${8.4}^\circ$ east at 110~m~a.s.l). 
This experiment studied extensive air showers in the primary energy range from $\sim 500$\,TeV to 100\,PeV.
The experiment collected data from different setups, but in this research, we use the data from the main \mbox KASCADE array only.
This array was composed of 252~scintillator detectors placed in a rectangular grid covering an area of~\mbox{$200 \times 200$\,m${}^2$}. 
The outer 192 detectors contained a shielding layer to detect electromagnetic and muon-dominated EAS parts separately.

The experimental and Monte~Carlo data we are using in this study were provided by the KCDC service~\cite{Haungs:2018xpw}.
Each event includes time-integrated deposits of electromagnetic and muon EAS components from the \mbox KASCADE array stations, as well as reconstructed features: primary energy~($E$), zenith angle~($\theta$), azimuthal angle~($\phi$), shower core position~($x$,~$y$), number of electrons~($N_e$) and muons~($N_\mu$), and shower age~($s$).
Details of description of these parameters can be found in Refs.~\cite{KCDC_manual, kgml_method}.
The values $N_e$, $N_\mu$, and $s$ are determined by fitting the lateral distribution function of the particle densities with the modified NKG function~\cite{KCDC_manual}. 
$E$ is reconstructed with the standard KASCADE algorithms by taking into account both $N_e$ and $N_\mu$ corrected for atmospheric attenuation depending on $\theta$, see Ref.~\cite{KCDC_manual} for more details. 
The resolution of $E$ is around {$11 \%$} in terms of the decimal logarithm of the ratio of the simulated to the reconstructed energies for MC events based on the \mbox{QGSJet-II.04} hadronic interaction model, which passed the quality cuts in the studied energy range~(see below)~\footnote{The difference from the energy resolution shown in our methodological paper~\cite{kgml_method} is due to usage of \mbox{QGSJet-II.04} hadronic model instead of \mbox{QGSJet-II.02} model~\cite{Ostapchenko:2004ss}.}.

The full efficiency of the trigger and reconstruction is reached at $E > 10^{15}$~eV.
We use the quality cuts recommended by KASCADE~\cite{kascade_cuts}: ${\theta < 18^\circ}$, ${\sqrt{x^2 + y^2} < 91}$~m, ${\log_{10} N_e > 4.8}$, ${\log_{10} N_\mu > 3.6}$, and the cut on the shower age set by KCDC: ${0.2 < s < 1.48}$~\cite{KCDC_manual}, which is stricter than the original one (${0.2 < s < 2.1}$). We also impose the additional cut on the reconstructed event energy $E > 10^{15.15}$~eV to ensure the stability of the unfolding procedure.

Overall, we have $\sim 3.5 \cdot 10^6$ experimental events in the studied energy range that passed the quality cuts.
The entire experimental dataset was divided into the so-called ``blind'' and ``unblind'' parts in an 80:20 ratio by random partitioning. The ``unblind'' part was used to check the correctness of the methods applied in our study~\cite{kgml_method}. The results for the ``blind'' part are disclosed in this paper. 
Also, we use a number of MC datasets provided by KCDC, that were simulated with CORSIKA~\cite{heck1998corsika} and accounted for the detector response for three different post-LHC hadronic interaction models: \mbox{QGSJet-II.04}~\cite{Ostapchenko:2010vb}, \mbox{EPOS-LHC}~\cite{Pierog:2013ria}, and \mbox{Sibyll~2.3c}~\cite{Riehn:2015aqb}. 
For low-energy hadronic interactions ($E < 200$~GeV), the FLUKA~\cite{Ferrari:2005zk} model has been used.
We have a total of $\sim 1.3 \cdot 10^5$, \mbox{$6.7 \cdot 10^4$}, and \mbox{$6.8 \cdot 10^4$} events that have passed the quality cuts in the studied energy range for the \mbox{QGSJet-II.04}, \mbox{EPOS-LHC} and \mbox Sibyll~2.3c models, respectively.
These MC sets include simulations for five primary particle types: protons~(\emph{p}), helium~(\emph{He}), carbon~(\emph{C}), silicon~(\emph{Si}), and iron~(\emph{Fe}).
In Ref.~\cite{kgml_method}, we used MC based on the \mbox{QGSJet-II.04} hadronic interaction model as a baseline,
in line with the original KASCADE research~\cite{Apel:2013uni} based on the \mbox{QGSJet-II.02} model~\cite{Ostapchenko:2004ss}.
The results for \mbox{EPOS-LHC} and \mbox{Sibyll~2.3c} are used for a calculation of the theoretical uncertainty.
\begin{figure}[t]
    \centering
    \includegraphics[width=\linewidth]{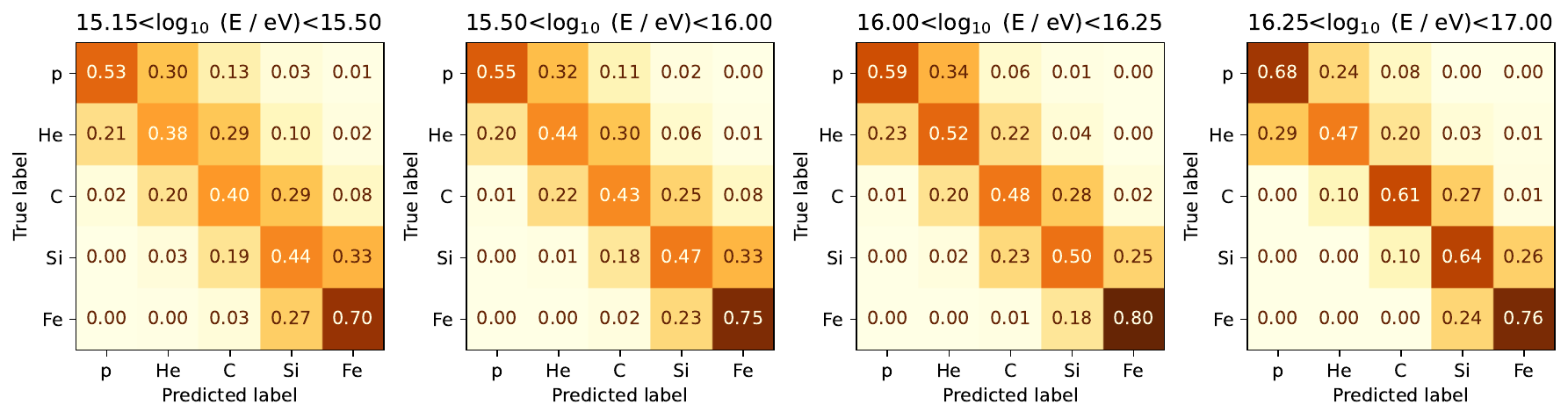}
    \caption{Confusion matrices of the CNN for the different reconstructed energy intervals. The neural network is trained and tested using the \mbox{QGSJet-II.04} hadronic interaction model.
    }
    \label{fig:confusion matrices energy bins}
\end{figure}

\section{Method and uncertainties}
\label{sec:method}
All details of our methods and their application are available in our methodological paper~\cite{kgml_method}.
In summary, we built a number of machine learning models to classify primary CRs into five groups: \emph{p}, \emph{He}, \emph{C}, \emph{Si} and \emph{Fe}, on an event-by-event basis. 
Namely, a random forest~(RF) classifier~\cite{random_forest}, a convolutional neural network~(CNN) inspired by the \mbox{LeNet-5} architecture~\cite{lenet_arch}, a multi-layer perceptron, and \mbox EfficientNetv2~\cite{tan2021efficientnetv2} were constructed.
These models were implemented using PyTorch~\cite{PyTorch}, TensorFlow~\cite{abadi2016tensorflow} and Scikit-learn~\cite{scikit-learn} packages.
The performance of these models was compared using their confusion matrices. 
This matrix represents the fraction of particles of each type that the given model classifies as this type (i.e., correctly) and all other types (i.e., incorrectly). 
The performance of the different classifiers was found to be almost similar; we selected the CNN for further use because of its robustness demonstrated in various other tests performed in our methodological paper~\cite{kgml_method}.
The CNN uses electromagnetic and muon deposits, $\theta$, $N_e$, $N_\mu$ and $s$ as input parameters.
An example of the CNN confusion matrix for different energy intervals is shown in Fig.~\ref{fig:confusion matrices energy bins}.

In Ref.~\cite{kgml_method}, we used the CNN to reconstruct the mass component spectra of CRs in the ``unblind'' part of the experimental data.
We enhanced the quality of mass component spectra reconstruction by performing the unfolding of our results with response matrices (that are transposed confusion matrices) derived from the MC simulations. 
We sequentially unfolded the primary energy and primary particle type using the Bayesian iterative unfolding procedure~\cite{DAgostini:1994fjx} with the pyunfold~\cite{pyunfold} Python package.
{In Ref.~\cite{kgml_method}}, all uncertainties related to particle type and energy reconstruction, as well as uncertainties of the unfolding procedure, were estimated using the MC set with the \mbox{QGSJet-II.02} hadronic interaction model and the ``unblind'' data set.
In what follows, we call these uncertainties ``basic'' ones. The results were shown to have better accuracy than that of the original KASCADE analysis~\cite{Apel:2013uni} (see, e.g. Fig.~20 of Ref.~\cite{kgml_method}).

In the present study, we apply the same analysis pipeline to the undisclosed portion (``blind'' part) of the experimental data. 
We perform the full unfolding and re-compute all the ``basic'' uncertainties
using the MC set based on the \mbox{QGSJet-II.04} hadronic model and the ``blind'' data set. Namely, the computed uncertainties are: effect of ``missing'' detectors in experimental data~({5 -- 18\,\%}, with respect to the reconstructed CR flux value), limited size of the MC set~({8 -- 25\,\%}), energy resolution for different MC mixtures~({13 -- 16\,\%}), spectral index in the MC set~({up to $4$\,\%}).
Another set of systematic biases comes from sequential energy and particle type unfoldings~({up to $8$\,\%}) and the unfolding procedure itself~({1 -- 24\,\%}).
All the uncertainty estimates shown in brackets above have been calculated as an average overall mass component for the studied energy range after the full unfolding. 
The estimates for two other hadronic interaction models considered in this study are almost similar (excluding outliers for \mbox{EPOS-LHC}, see below).
We calculated the total ``basic'' systematic uncertainty as the square root of the sum of the squares of the uncertainties and biases described above, conservatively considering that they are independent.
Compared to our analysis in Ref.~\cite{kgml_method}, we have applied a correction for detection efficiency to the computations. 
For example, for \mbox{QGSJet-II.04} hadronic model, the correction in the first energy bin~($6.15 < \log_{10}{(E/\text{GeV})} < 6.2425$) varies from $6\%$ for protons to $34\%$ for iron; in the second bin~($6.2425 < \log_{10}{(E/\text{GeV})} < 6.335$), it varies from $1\%$ to $8\%$, respectively, while in higher bins full efficiency is reached and no correction is applied.
Therefore, the accuracy of the spectra reconstruction in the two lower energy bins was improved, while the respective uncertainty vanished in the present analysis.

\paragraph{Theoretical uncertainty.}
To consider uncertainties associated with a difference between hadronic interaction models, we first perform the unfolding procedures for each model and compute the respective uncertainty bands of ``basic'' systematics.
This implies unfolding with the response matrices derived from the corresponding MC sets for our CNN.
We set the results obtained for the \mbox{QGSJet-II.04} model as our baseline, following the usage of the previous version of this model (\mbox{QGSJet-II.02}) in the original KASCADE mass composition study~\cite{Apel:2013uni}.
For each mass component spectrum in each energy bin, we
estimate the theoretical uncertainty as a range between the minimum and the maximum edges of the ``basic'' systematic uncertainty bands among all the hadronic models used.
In any case, we should note that all the considered hadronic models may be incorrect, so that true values of the mass component's fluxes may not be covered even by the theoretical uncertainties considered.

\paragraph{Mean logarithmic mass representation.}
The results of two other modern experiments, IceTop~\cite{IceCube:2019hmk} and TALE~\cite{TelescopeArray:2020bfv}, on the CR mass composition were presented in the form of $\left<\ln{A} \right>$ depending on the energy.
Therefore, to make a reasonable comparison, we compute this quantity in the present paper too.
We derive $\left<\ln{A} \right>$ by taking the weighted sum of the fluxes of the mass component spectra in each energy bin.
To estimate the ``basic'' systematic uncertainty for $\left<\ln{A} \right>$, we first calculate separately each uncertainty contribution of those discussed above.
The calculation is performed using the propagation of uncertainty method.
The contributions of the unfolding biases are calculated as the direct weighted sum of the biases. 
For the ``basic'' systematic uncertainty, we combine all the contributions in quadrature, considering them to be uncorrelated.
To account for the theoretical uncertainty of the hadronic models, we use the same method as for the spectra of individual mass components, namely we calculate $\left<\ln{A} \right>$ for each hadronic interaction model {separately, including the ``basic'' systematic uncertainties}. 
We take the result for the \mbox{QGSJet-II.04} model as a baseline and
{estimate the theoretical uncertainty}
as a range between the lowest and highest edges of the ``basic'' uncertainty bands among all the hadronic models used.
\begin{figure}[t]
    \centering
    \includegraphics[width=\textwidth]{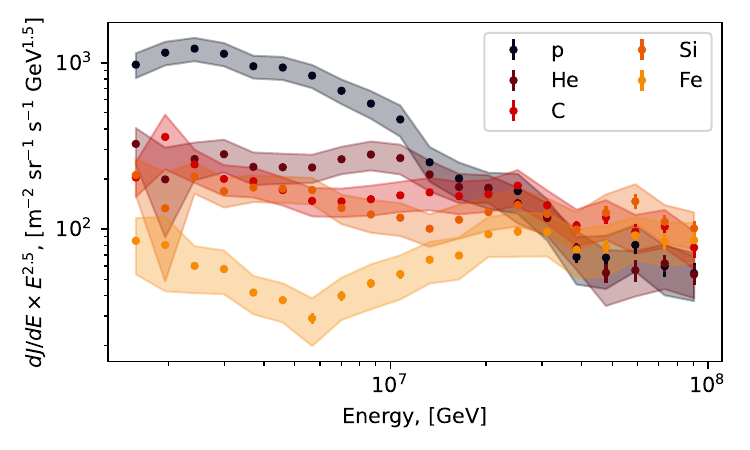}
    \caption{
    Mass component spectra reconstructed with our CNN method for the ``blind'' part of the experimental data using \mbox QGSJet-II.04 hadronic interaction model and the full unfolding procedure.
    Error bars display the statistical uncertainties, while bands represent the ``basic'' systematic uncertainties.
    }
    \label{fig:spectra qgs04}
\end{figure}

\section{Analysis and results}
\label{sec:results}
\begin{figure}[t]
    \centering
    \includegraphics[width=\textwidth]{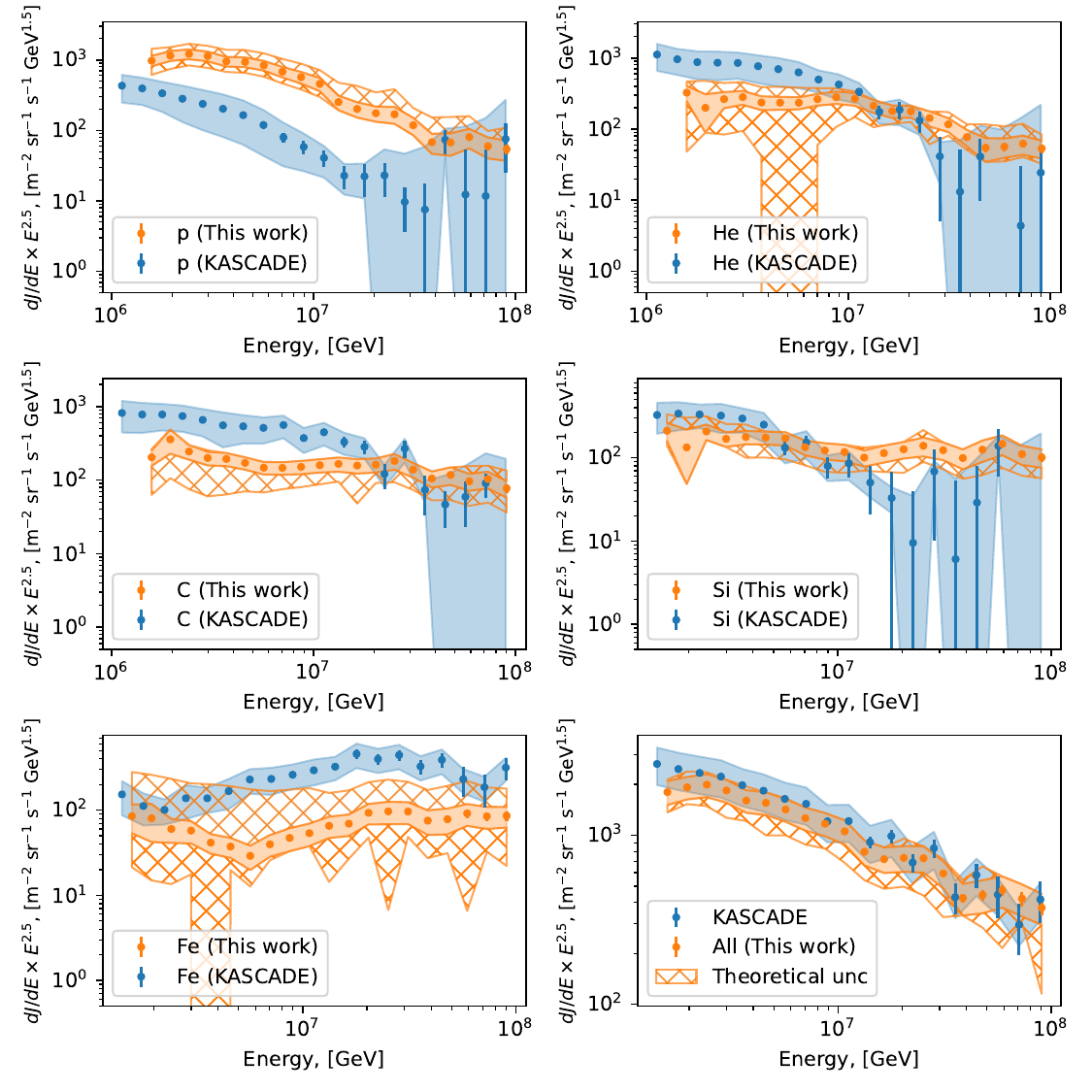}
    \caption{
    Comparison of the individual mass component spectra and all-particle spectra~(right bottom) of the original KASCADE analysis~\cite{Apel:2013uni} (in blue) and our CNN analysis for the ``blind'' part of the experimental data~(in orange).
    {Original KASCADE results are obtained for \mbox QGSJet-II.02 hadronic interaction model, CNN results are for QGSJet-II.04 model.
    Error bars display the statistical uncertainties, solid bands represent the ``basic'' systematic uncertainties, hatched bands show the estimation of theoretical uncertainties related to post-LHC hadronic models. The latter is not computed for original KASCADE results.}
    }
    \label{fig:spectra final}
\end{figure}
In this section, we show the energy spectra for elementary groups of cosmic rays obtained with the ``blind'' part of the KASCADE experimental data.
The spectra reconstructed using \mbox{QGSJet-II.04} hadronic interaction model are presented in Fig.~\ref{fig:spectra qgs04}.
Here, we include all the ``basic'' uncertainties described in Section~\ref{sec:method},
they have values in a range $(20 - 45) \%$ 
depending on the energy, when averaged over all mass components.

The comparison between the original KASCADE mass component spectra (based on the \mbox{QGSJet-II.02} hadronic model) and our final CNN results~(based on the \mbox{QGSJet-II.04} hadronic model) is shown in Fig.~\ref{fig:spectra final}~\footnote{This figure resembles the Fig.~20 from our methodological paper~\cite{kgml_method}, but they are quite different. 
Here for our CNN results, we use the data from the ``blind'' part of the experimental dataset 
and \mbox{QGSJet-II.04} hadronic model as a baseline, instead of ``unblind'' dataset and \mbox{QGSJet-II.02} model used in that study. Also, the theoretical uncertainty of hadronic models was not estimated in our methodological study. The results of the original KASCADE experiment are the same in both figures.
}. 
We should note, that the differences in the spectra of each component are larger than the respective differences estimated in our methodological paper~\cite{kgml_method}, where both analysis methods use one and the same QGSJet-II.02 hadronic model.
{For our results we also show separately the estimation of the theoretical uncertainties related to hadronic models,
while there is no such an estimation for the original KASCADE results.}
{We estimate the magnitude of the theoretical uncertainties, defining it as a
relative difference between the central values of the flux and the most distant edge of the theoretical uncertainty bands  (unlike ``basic'' systematics, the theoretical uncertainty band is non-symmetric).
We get a value in a range of $(70 - 150)\%$, depending on the energy, when averaged over all mass components.}

One can see that within the uncertainties, the all-particle flux is in agreement between the original KASCADE and our CNN results.
The statistical uncertainties of the \mbox{KASCADE} spectra exceed those of the CNN spectra because the experimental dataset provided by KCDC, which we use in this study, is larger by a factor of about \textit{eleven} compared to the dataset used in the KASCADE study~\cite{Apel:2013uni}~(not all of the experimental data, that are now available via KCDC, was used in the original research). 
At the same time, the uncertainties are dominated by systematic ones for almost all mass components at all energies for both analyses.
As one can see, the ``basic'' systematic uncertainties of the CNN analysis are smaller than those of the original \mbox{KASCADE} analysis, both for individual components and for all-particle spectra, which is an achievement of our method, while the theoretical uncertainties of the CNN are generally comparable to them.
Also, our estimation of the theoretical uncertainties yields outliers at several energies (for \emph{He} and \emph{Fe} components).
This effect is entirely due to the estimation of the unfolding bias uncertainty for the \mbox EPOS-LHC hadronic model. 
In any case, the most pronounced discrepancy between the original KASCADE and the CNN results is the excess of the \emph{p} component at low and intermediate energies in the CNN analysis. We should stress, that this discrepancy is seen in all the hadronic interaction models we consider (including the same model QGSJet-II.02, tested in our Ref.~\cite{kgml_method}), rendering difficult its explanation by the difference in the models.
\begin{figure}[t]
    \captionsetup[subfigure]{aboveskip=0pt,belowskip=-2pt}
    \centering
    \begin{subfigure}{.48\textwidth}
        \includegraphics[width=\textwidth]{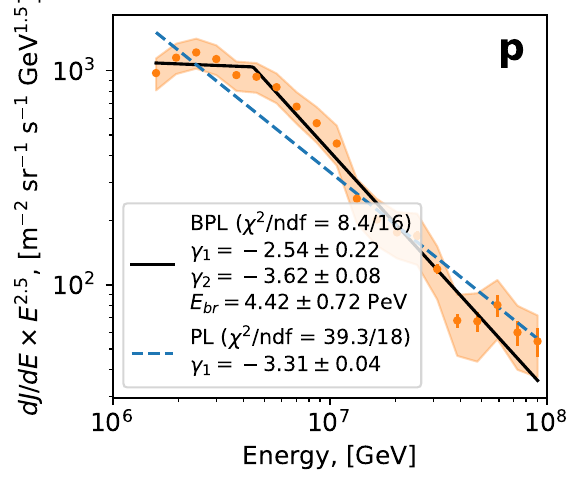}
       \caption{}
    \end{subfigure}
    \hfill
    \begin{subfigure}{.48\textwidth}
        \includegraphics[width=\textwidth]{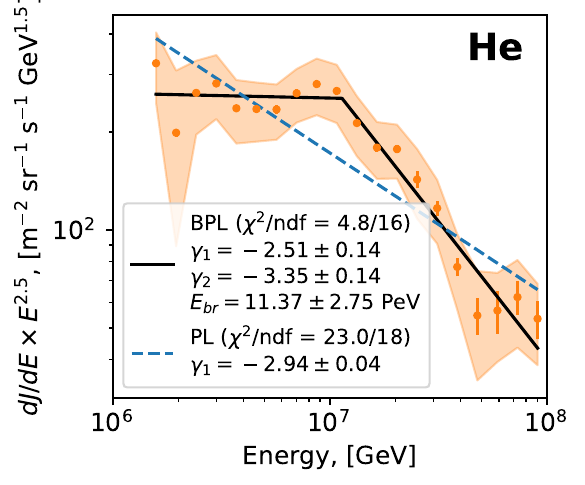}
       \caption{}
    \end{subfigure}
    \vfill
    \begin{subfigure}{.8\textwidth}
        \includegraphics[width=\textwidth]{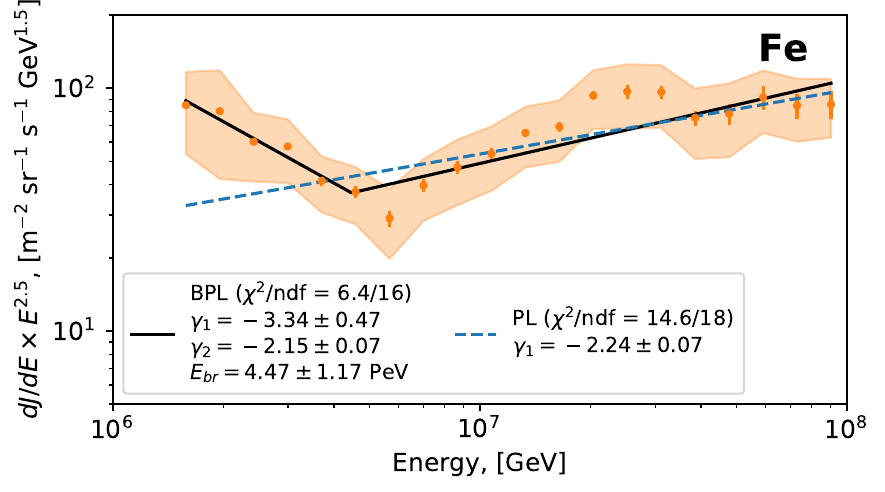}
       \caption{}
    \end{subfigure}
    \caption{
    Individual spectra for (a)~protons, (b)~helium, and (c)~iron 
    fitted with PL~(see Eq.~\ref{eq:PL}) and BPL~(see Eq.~\ref{eq:BPL}), represented by the blue dashed line and black solid line, respectively.
    Error bars display the statistical uncertainties, while bands represent the ``basic'' systematic uncertainties.
    Fit qualities and results are depicted in the legends of the corresponding figures.
    }
    \label{fig:spectrum fits}
\end{figure}

\paragraph{Fits for the spectra of individual mass components.}
We are searching for breaks in the obtained individual spectra.
The spectra are fitted using a simple power law~(PL):
\begin{equation}\label{eq:PL}
    dJ/dE = J_0 \cdot E^{\gamma_1},
\end{equation}
where $J_0$ is a normalization factor and $\gamma_1$ is a spectral index,
and a broken power law~(BPL):
\begin{equation}\label{eq:BPL}
    dJ/dE = J_0 
    \begin{cases}
      \left(E/E_{\rm br}\right)^{\gamma_1}, & \text{if}\ E < E_{\rm br} \\
      \left(E/E_{\rm br}\right)^{\gamma_2}, & \text{otherwise},
    \end{cases}
\end{equation}
where $E_{\rm br}$ is the energy of the break and $\{\gamma_1, \gamma_2\}$ are the spectral indices before and after the break.
The $\chi^2$ fit calculations are performed with the iminuit~\cite{JAMES1975343, dembinski_2024_10638795} package.
The significance of the break is estimated by comparing PL and BPL fits, considering both statistical and ``basic'' systematic uncertainties summed in quadrature.
Fig.~\ref{fig:spectrum fits} shows the mass component spectra that have a statistically significant deviation from the PL. The numerical results of the fits are also shown in this figure. 
In the \textit{p}~spectrum, the use of the BPL reduces 
$\chi^2$ by 30.9 at the expense of 2 additional degrees of freedom~(d.o.f.),
this results in a p-value of~$1.96 \cdot 10^{-7}$ ($5.2\sigma$, two-sided). 
In the case of the \textit{He}~spectrum the similar value of $\Delta\chi^2$ is 18.2 and the resulting p-value is~$1.07 \cdot 10^{-4}$ ($3.9\sigma$). Finally, for \textit{Fe} BPL reduces $\chi^2$ by 8.2 and the p-value is $1.62\cdot10^{-2}$ ($2.4\sigma$). As one can see, the breaks for \textit{p} and \textit{He} are knee-like, while for~\textit{Fe} the spectrum exhibits a hardening.
Other individual mass component~(\textit{C}, \textit{Si}) spectra show no significant deviations from the PL within the energy range considered. The values of PL spectral indices are $\gamma_1 = -2.71 \pm 0.04$ with $\chi^2/{\rm d.o.f.} = 0.42$ and $\gamma_1 = -2.67 \pm 0.05$ with $\chi^2/{\rm d.o.f.} = 0.41$ for \textit{C} and \textit{Si} respectively.
The PL fit for all-particle spectrum yields $\gamma_1 = -2.95 \pm 0.03$ with $\chi^2/{\rm d.o.f.} = 0.43$.
When fitted with BPL using the same method (including basic systematic uncertainties), the all-particle spectrum exhibits a knee at around 3~PeV but its significance with respect to the PL fit is low, similar to the original KASCADE result~\cite{kascade_cuts}. 

\begin{figure}[t]
    \centering
    \includegraphics[width=\textwidth]{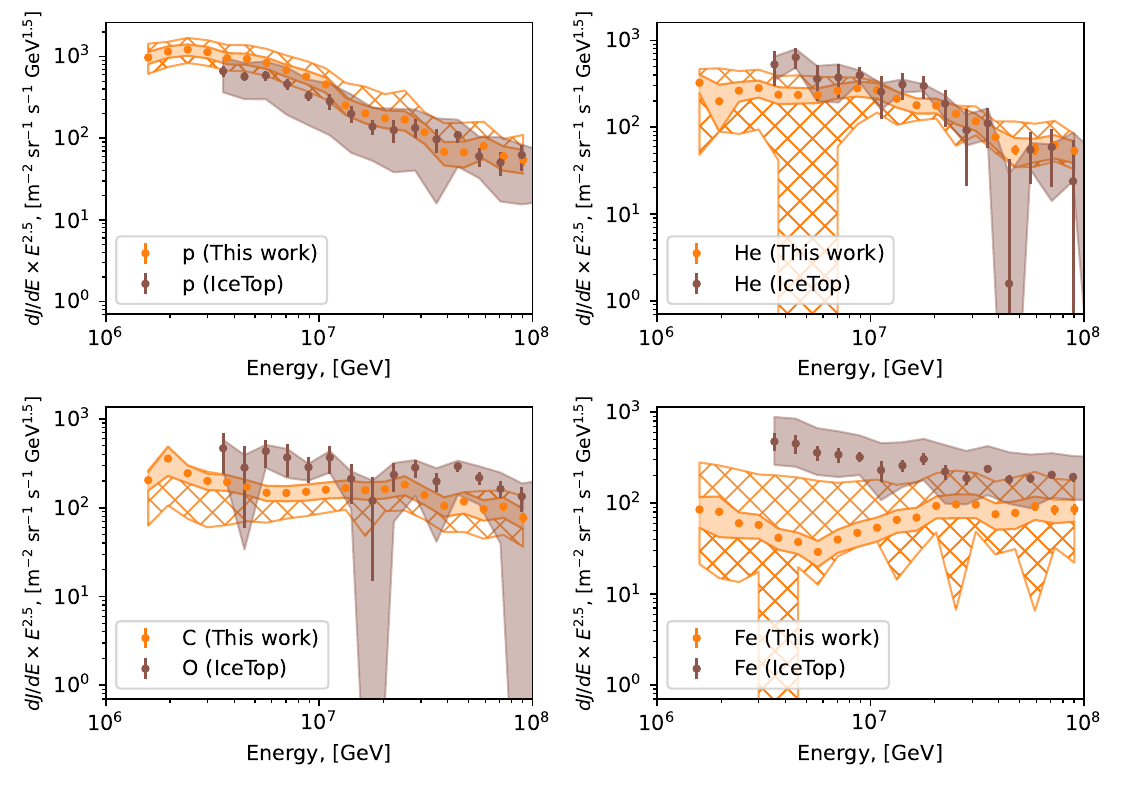}
    \caption{
    Comparison of the mass component spectra of IceTop~\cite{IceCube:2019hmk}
    (in brown) and our CNN spectra for the ``blind'' part of the experimental data~(in orange).
    IceTop results are obtained for \mbox Sibyll~2.1 hadronic interaction model, CNN results are for the QGSJet-II.04 model.
    Error bars display the statistical uncertainties, solid bands represent the ``basic'' systematic uncertainties, hatched bands show the estimation of theoretical uncertainties related to hadronic models. The latter is not computed for IceTop results.
    }
    \label{fig:spectra final comp icetop}
\end{figure}

\paragraph{Comparison with the results of other experiments.} We also compare our CNN mass component spectra with the IceTop results~\cite{IceCube:2019hmk} in Fig.~\ref{fig:spectra final comp icetop}.
Note that the IceTop results are obtained using the \mbox Sibyll~2.1 hadronic interaction model~\cite{Ahn:2009wx}, while the theoretical uncertainty associated with a difference between hadronic models is taken into account in a different way than ours. Therefore, we do not show the theoretical uncertainty of the IceTop results in this comparison.
They also consider four different mass components: \emph{p}, \emph{He}, \emph{O}, and \emph{Fe}, instead of the five components considered in our analysis.
Therefore, the comparison is not ideal, however, one can see a general agreement between these results within the uncertainties.

\begin{figure}[htbp!]
    \centering
    \includegraphics[width=0.9\textwidth]{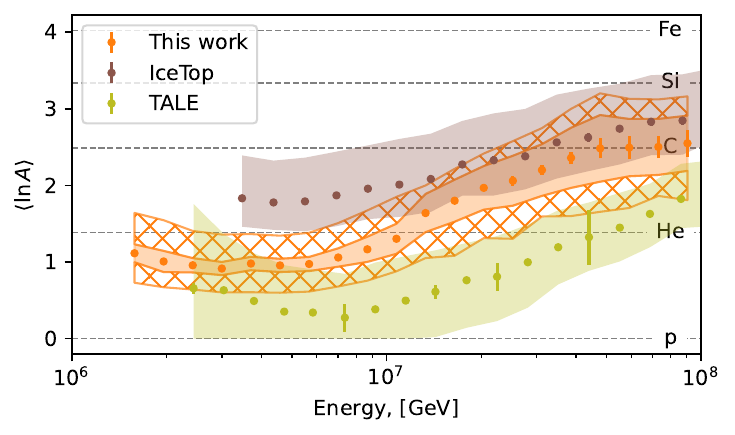}
    \vfill
    \includegraphics[width=0.9\textwidth]{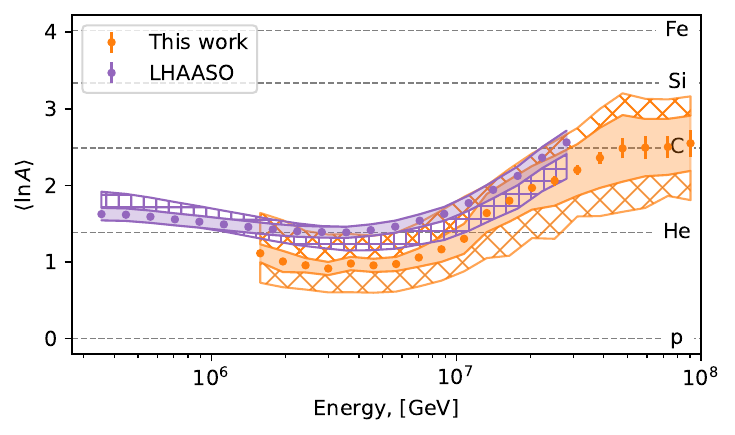}
    \caption{
    Comparison of the mean logarithmic mass ($\left<\ln{A} \right>$) of cosmic-ray flux for different experiments. 
    Error bars display the statistical uncertainties, solid bands represent the ``basic'' systematic uncertainties, hatched bands show the estimation of theoretical uncertainties related to hadronic models.
    Our CNN analysis for the ``blind'' part of the experimental data shown in orange, based on \mbox{QGSJet-II.04}.
    \textit{Top panel:} comparison with \mbox{IceTop}~\cite{IceCube:2019hmk}
    (in brown) and \mbox{TALE}~\cite{TelescopeArray:2020bfv}
    (in green). 
    {IceTop results are obtained for \mbox Sibyll~2.1 model, TALE results are for EPOS-LHC model.}
    \textit{Bottom panel:} comparison with \mbox{LHAASO}~\cite{LHAASO:2024knt} (in purple)
    {, that is using \mbox{QGSJet-II.04} model for basic systematic uncertainties}.
    Theoretical uncertainties are not computed for IceTop and TALE results. For \mbox{LHAASO} they are covering \mbox{QGSJet-II.04}, \mbox{EPOS-LHC} and \mbox{Sibyll 2.3d} models;
    for this work: \mbox{QGSJet-II.04}, \mbox{EPOS-LHC} and \mbox{Sibyll 2.3c} models.
    }
    \label{fig:lnA final comp}
\end{figure}

Finally, the comparison of the results from our CNN method, IceTop~\cite{IceCube:2019hmk}, TALE~\cite{TelescopeArray:2020bfv} and the recent result of LHAASO~\cite{LHAASO:2024knt} in terms of $\left<\ln{A} \right>$ is shown in Fig.~\ref{fig:lnA final comp}.
In our method $\left<\ln{A} \right>$ is calculated as described in Section~\ref{sec:method}.
Our result is based on the \mbox{QGSJet-II.04} hadronic interaction model and takes into account the theoretical uncertainty related to the difference between hadronic models.

We should note that the results based on the \mbox{QGSJet-II.04} hadronic interaction model are in the middle of the theoretical uncertainty band covered also by the \mbox{EPOS-LHC} and \mbox{Sibyll~2.3c} models.
The \mbox{EPOS-LHC} model predicts a lighter composition, while \mbox{Sibyll~2.3c} is predicting a heavier one.
It is interesting to notice, that the result for each hadronic model tends towards the results of another experiment that uses the same model or at least a model from the same family.
Namely, the top edge of our theoretical uncertainties is closer to the
results of IceTop that are based on \mbox{Sibyll~2.1} model, while our bottom edge --- to the TALE results that are based on \mbox{EPOS-LHC} model~\footnote{We do not show the estimates of the theoretical uncertainty for these two experiments because they were performed with a different approach than ours and are therefore difficult to compare with our estimate.}.
This indicates that at least a part of the discrepancy between the results of the different experiments can be explained by the difference in the hadronic interaction models they use.

LHAASO experiment provides the results of its analysis~\cite{LHAASO:2024knt} separately for each hadronic interaction model: QGSJet-II.04, EPOS-LHC, and Sibyll 2.3d~\cite{Riehn:2019jet}. 
We compare the results of LHAASO with our results directly, including the theoretical uncertainty due to hadronic models inherent to each result. 
One can see that the results exhibit good agreement within the theoretical uncertainties, while the uncertainty itself is lower in the LHAASO measurement.
This level of agreement looks satisfactory, given the differences between the experiments.

The numerical tables of our results are presented in Appendix~\ref{sec:app}.

\section{Discussion and conclusion}
\label{sec:conclusion}
In this research we have presented the cosmic-ray mass component spectra reconstructed with the specific convolutional neural network method using the
KASCADE experiment archival data. We have performed the full unfolding analysis and computed all uncertainties affecting the reconstruction and unfolding procedures. Also, we have estimated the theoretical uncertainty arising due to differences between post-LHC hadronic interaction models: \mbox{QGSJet-II.04}, \mbox{EPOS-LHC} and \mbox{Sibyll~2.3c}.
We compared our mass component spectra with the original KASCADE results~\cite{Apel:2013uni} and with IceTop results~\cite{IceCube:2019hmk}. Our all-particle spectrum is consistent with the original KASCADE one. However,
{the observed proton flux is higher than that of the original KASCADE in almost all energy range studied, up to a factor of {$\sim 10$}
at $E = 10$~PeV, in terms of differential flux}.
At the same time, the individual mass components spectra are mostly in agreement with those of IceTop.  
We also converted our results to a form of $\left<\ln{A} \right>$ and compared them with those obtained by IceTop~\cite{IceCube:2019hmk}, TALE~\cite{TelescopeArray:2020bfv}, and LHAASO~\cite{LHAASO:2024knt}.
Our CNN results are generally in agreement with the three mentioned experiments if we take into account the theoretical uncertainties related to the difference between the hadronic interaction models.
The ``basic'' systematic uncertainties of our results are smaller than those of IceTop and TALE results, and comparable to those of LHAASO results, over the most part of the energy range studied.
Finally, we found that the theoretical uncertainty is the dominant one in our analysis. However, it is comparable to the uncertainties of the IceTop and TALE results, that do not take into account the difference between the hadronic models.

In the case of the ``basic'' uncertainties, 
we have estimated that the significant contributions come from the energy resolution of the different MC mixtures and from the bias of the sequential energy and particle type unfoldings.
This opens up the possibility of reducing the ``basic'' uncertainties of the analysis 
by using a more realistic (in terms of mass components) and larger Monte Carlo set to accurately account for an energy resolution in different mixtures, and by incorporating the correction for the bias of the sequential unfolding procedure.
At the same time the largest uncertainty of the whole analysis is the theoretical uncertainty related to the difference between the hadronic interaction models.
This one can only be reduced by further understanding of the EAS physics and respective improvement of hadronic models.

We also performed fits of the individual components spectra with power-law and broken power-law functions to identify possible spectral features.
We have found that the spectra of the proton and helium components show knee-like features at energies around 4.4~PeV and 11~PeV, respectively.
However, the carbon and silicon components do not show any 
signs of spectral breaks.
While the spectrum of the iron component
shows a hint~($2.4\sigma$) of the break at $\sim 4.5$~PeV with a spectrum hardening above this energy.
The knee-like feature in protons was seen by KASCADE and TALE experiments~\cite{kascade_cuts, Apel:2013uni, TelescopeArray:2020bfv} and a feature in helium was also seen in the original KASCADE analysis~\cite{kascade_cuts, Apel:2013uni}, though at a somewhat lower energy. In the present study we have shown, for the first time, that these features have high statistical significance; we also computed the respective spectral indices.
The positions of both features (within their uncertainties) are consistent with their rigidity dependent origin. This confirms the theory that the knee in the all-particle energy spectrum appears due to the respective knee in the proton spectrum. 
Although not very significant, the hardening in the iron spectrum at this energy was never seen before.
It is remarkable that the energy of this break corresponds to that of the break in the proton spectrum at $166$~TeV recently observed by the \mbox{GRAPES-3} experiment~\cite{PhysRevLett.132.051002}, if one would assume the rigidity dependent origin of both brakes.
We think that this {finding} deserves a further study.

\acknowledgments
We are grateful to Dmitry Kostunin for the idea of this study and his valuable assistance at all stages of this research.
We also would like to thank
Grigory Rubtsov, Ivan Kharuk, and Vladimir Lenok for rich discussions, and Victoria Tokareva for the help in the preparation of the KCDC data. In addition, we would like to thank the entire KCDC team for the high-quality data and extra MC.
We also thank the anonymous referee for the idea of fitting the spectra obtained.
The work was supported by the Russian Science Foundation grant \mbox 22-22-00883. 

\appendix
\section{Table of results}
\label{sec:app}
The values of $\left<\ln{A} \right>$ for our CNN analysis are shown in Fig.~\ref{fig:lnA final comp} are represented in Table~\ref{tab:ln A combined}.
The individual mass component spectra and the all-particle spectrum shown in Fig.~\ref{fig:spectra final} are represented in Table~\ref{tab:mass component fluxes} and Table~\ref{tab:total flux}, respectively.
\begin{table}[htbp!]
    \centering
    \caption{
    $\left<\ln{A} \right>$ values for our CNN analysis. 
    ``$\pm$\,Stat'' means statistical uncertainties. 
    ``$\pm$\,Bas.\,sys'' means the ``basic'' systematic uncertainties for \mbox{QGSJet-II.04} model. 
    ``Th.\,unc.'' means the theoretical uncertainty band, 
    ``$+$''~is the upper uncertainty and ``$-$''~is the lower uncertainty.
    The theoretical uncertainty covers results based on \mbox{QGSJet-II.04}, \mbox{EPOS-LHC}, and \mbox{Sibyll 2.3c}.
    }
    \label{tab:ln A combined}
    \smallskip
    \begin{tabular}{|l|ccccc|}
    \hline
    \makecell{ Energy bin, \\ $\log_{10}{(E/\text{GeV})}$ } & $\left<\ln{A} \right>$ & $\pm$\,Stat & $\pm$\,Bas.\,sys & \multicolumn{2}{c|}{$-$ Th.\,unc $+$} \\
    \hline
    $ 6.15 - 6.2425 $  &  1.114 & 0.008 & 0.117  & 0.385 & 0.526 \\
    $ 6.2425 - 6.335 $ & 1.007  & 0.013 & 0.138  & 0.336 & 0.538 \\
    $ 6.335 - 6.4275 $ & 0.957  & 0.011 & 0.094  & 0.317 & 0.448 \\
    $ 6.4275 - 6.52 $  &  0.914 & 0.011 & 0.086  & 0.311 & 0.443 \\
    $ 6.52 - 6.6125 $  &  0.980 & 0.013 & 0.088  & 0.374 & 0.384 \\
    $ 6.6125 - 6.705 $ & 0.955  & 0.015 & 0.087  & 0.357 & 0.388 \\
    $ 6.705 - 6.7975 $ & 0.973  & 0.017 & 0.092  & 0.360 & 0.408 \\
    $ 6.7975 - 6.89 $  &  1.058 & 0.020 & 0.120  & 0.378 & 0.450 \\
    $ 6.89 - 6.9825 $  &  1.165 & 0.023 & 0.164  & 0.409 & 0.507 \\
    $ 6.9825 - 7.075 $ & 1.303  & 0.027 & 0.196  & 0.425 & 0.547 \\
    $ 7.075 - 7.1675 $ & 1.638  & 0.025 & 0.249  & 0.587 & 0.360 \\
    $ 7.1675 - 7.26 $  &  1.799 & 0.034 & 0.274  & 0.718 & 0.416 \\
    $ 7.26 - 7.3525 $  &  1.967 & 0.038 & 0.285  & 0.656 & 0.448 \\
    $ 7.3525 - 7.445 $ & 2.060  & 0.064 & 0.326  & 0.757 & 0.517 \\
    $ 7.445 - 7.5375 $ & 2.200  & 0.064 & 0.338  & 0.606 & 0.547 \\
    $ 7.5375 - 7.63 $  &  2.359 & 0.082 & 0.393  & 0.760 & 0.639 \\
    $ 7.63 - 7.7225 $  &  2.482 & 0.131 & 0.435  & 0.828 & 0.718 \\
    $ 7.7225 - 7.815 $ & 2.492  & 0.155 & 0.374  & 0.787 & 0.637 \\
    $ 7.815 - 7.9075 $ & 2.502  & 0.148 & 0.366  & 0.639 & 0.619 \\
    $ 7.9075 - 8.0 $   &  2.549 & 0.174 & 0.360  & 0.741 & 0.613 \\
    \hline
    \end{tabular}
\end{table}
\begin{table}[htbp!]
\footnotesize
\centering
\caption{Individual mass component differential flux~($dJ/dE$) in form:
$dJ/dE$ $\pm$ statistical uncertainty $\pm$ ``basic'' systematic uncertainty (based on \mbox{QGSJet-II.04} model), $\left[ \text{m}^{-2} \text{sr}^{-1} \text{s}^{-1} \text{GeV}^{-1} \right]$.}
\smallskip
\label{tab:mass component fluxes}
\begin{tabular}{|l|ccc|}
\hline
\makecell{ Energy bin, \\ $\log_{10}{(E/\text{GeV})}$ } & \textit{p} flux & \textit{He} flux & \textit{C} flux \\
\hline
$6.15-6.2425$ & $( 3.105 \pm 0.010 \pm 0.528 ) \times 10^{-13}$ & $( 1.037 \pm 0.012 \pm 0.253 ) \times 10^{-13}$ & $( 6.513 \pm 0.064 \pm 1.544 ) \times 10^{-14}$ \\
$6.2425-6.335$ & $( 2.152 \pm 0.007 \pm 0.348 ) \times 10^{-13}$ & $( 3.724 \pm 0.099 \pm 2.056 ) \times 10^{-14}$ & $( 6.701 \pm 0.128 \pm 2.424 ) \times 10^{-14}$ \\
$6.335-6.4275$ & $( 1.336 \pm 0.005 \pm 0.214 ) \times 10^{-13}$ & $( 2.898 \pm 0.060 \pm 0.742 ) \times 10^{-14}$ & $( 2.691 \pm 0.050 \pm 0.607 ) \times 10^{-14}$ \\
$6.4275-6.52$ & $( 7.307 \pm 0.034 \pm 1.164 ) \times 10^{-14}$ & $( 1.819 \pm 0.038 \pm 0.405 ) \times 10^{-14}$ & $( 1.293 \pm 0.026 \pm 0.270 ) \times 10^{-14}$ \\
$6.52-6.6125$ & $( 3.610 \pm 0.019 \pm 0.565 ) \times 10^{-14}$ & $( 8.963 \pm 0.211 \pm 1.981 ) \times 10^{-15}$ & $( 7.357 \pm 0.172 \pm 1.493 ) \times 10^{-15}$ \\
$6.6125-6.705$ & $( 2.082 \pm 0.013 \pm 0.329 ) \times 10^{-14}$ & $( 5.234 \pm 0.138 \pm 1.079 ) \times 10^{-15}$ & $( 3.814 \pm 0.103 \pm 0.735 ) \times 10^{-15}$ \\
$6.705-6.7975$ & $( 1.093 \pm 0.009 \pm 0.174 ) \times 10^{-14}$ & $( 3.064 \pm 0.092 \pm 0.591 ) \times 10^{-15}$ & $( 1.930 \pm 0.063 \pm 0.375 ) \times 10^{-15}$ \\
$6.7975-6.89$ & $( 5.204 \pm 0.053 \pm 0.878 ) \times 10^{-15}$ & $( 2.017 \pm 0.058 \pm 0.378 ) \times 10^{-15}$ & $( 1.125 \pm 0.036 \pm 0.218 ) \times 10^{-15}$ \\
$6.89-6.9825$ & $( 2.558 \pm 0.034 \pm 0.484 ) \times 10^{-15}$ & $( 1.264 \pm 0.038 \pm 0.250 ) \times 10^{-15}$ & $( 6.811 \pm 0.224 \pm 1.399 ) \times 10^{-16}$ \\
$6.9825-7.075$ & $( 1.207 \pm 0.020 \pm 0.256 ) \times 10^{-15}$ & $( 7.069 \pm 0.221 \pm 1.438 ) \times 10^{-16}$ & $( 4.218 \pm 0.135 \pm 0.869 ) \times 10^{-16}$ \\
$7.075-7.1675$ & $( 3.916 \pm 0.080 \pm 0.920 ) \times 10^{-16}$ & $( 3.308 \pm 0.077 \pm 0.693 ) \times 10^{-16}$ & $( 2.580 \pm 0.066 \pm 0.568 ) \times 10^{-16}$ \\
$7.1675-7.26$ & $( 1.837 \pm 0.050 \pm 0.454 ) \times 10^{-16}$ & $( 1.633 \pm 0.048 \pm 0.324 ) \times 10^{-16}$ & $( 1.437 \pm 0.045 \pm 0.319 ) \times 10^{-16}$ \\
$7.26-7.3525$ & $( 9.385 \pm 0.302 \pm 2.289 ) \times 10^{-17}$ & $( 9.485 \pm 0.281 \pm 1.789 ) \times 10^{-17}$ & $( 8.669 \pm 0.288 \pm 1.881 ) \times 10^{-17}$ \\
$7.3525-7.445$ & $( 5.316 \pm 0.257 \pm 1.430 ) \times 10^{-17}$ & $( 4.490 \pm 0.276 \pm 1.085 ) \times 10^{-17}$ & $( 5.732 \pm 0.287 \pm 1.386 ) \times 10^{-17}$ \\
$7.445-7.5375$ & $( 2.181 \pm 0.118 \pm 0.607 ) \times 10^{-17}$ & $( 2.152 \pm 0.112 \pm 0.478 ) \times 10^{-17}$ & $( 2.570 \pm 0.130 \pm 0.583 ) \times 10^{-17}$ \\
$7.5375-7.63$ & $( 7.364 \pm 0.546 \pm 2.333 ) \times 10^{-18}$ & $( 8.341 \pm 0.546 \pm 2.179 ) \times 10^{-18}$ & $( 1.138 \pm 0.069 \pm 0.281 ) \times 10^{-17}$ \\
$7.63-7.7225$ & $( 4.285 \pm 0.458 \pm 1.510 ) \times 10^{-18}$ & $( 3.475 \pm 0.445 \pm 1.281 ) \times 10^{-18}$ & $( 7.481 \pm 0.650 \pm 2.013 ) \times 10^{-18}$ \\
$7.7225-7.815$ & $( 3.000 \pm 0.356 \pm 0.925 ) \times 10^{-18}$ & $( 2.112 \pm 0.317 \pm 0.644 ) \times 10^{-18}$ & $( 3.607 \pm 0.420 \pm 0.939 ) \times 10^{-18}$ \\
$7.815-7.9075$ & $( 1.312 \pm 0.178 \pm 0.434 ) \times 10^{-18}$ & $( 1.364 \pm 0.161 \pm 0.410 ) \times 10^{-18}$ & $( 2.271 \pm 0.237 \pm 0.585 ) \times 10^{-18}$ \\
$7.9075-8.0$ & $( 6.995 \pm 1.096 \pm 2.247 ) \times 10^{-19}$ & $( 6.873 \pm 0.926 \pm 1.927 ) \times 10^{-19}$ & $( 9.971 \pm 1.282 \pm 2.534 ) \times 10^{-19}$ \\
\hline
 & \textit{Si} flux & \textit{Fe} flux &   \\
\hline
$6.15-6.2425$    & $( 6.735 \pm 0.089 \pm 1.654 ) \times 10^{-14}$ & $( 2.711 \pm 0.059 \pm 1.007 ) \times 10^{-14}$ & \\
$6.2425-6.335$   & $( 2.497 \pm 0.078 \pm 1.592 ) \times 10^{-14}$ & $( 1.503 \pm 0.043 \pm 0.712 ) \times 10^{-14}$ & \\
$6.335-6.4275$   & $( 2.276 \pm 0.047 \pm 0.481 ) \times 10^{-14}$ & $( 6.606 \pm 0.217 \pm 2.079 ) \times 10^{-15}$ & \\
$6.4275-6.52$    & $( 1.091 \pm 0.027 \pm 0.221 ) \times 10^{-14}$ & $( 3.711 \pm 0.128 \pm 1.089 ) \times 10^{-15}$ & \\
$6.52-6.6125$    & $( 6.738 \pm 0.157 \pm 1.234 ) \times 10^{-15}$ & $( 1.573 \pm 0.073 \pm 0.408 ) \times 10^{-15}$ & \\
$6.6125-6.705$   & $( 3.893 \pm 0.104 \pm 0.696 ) \times 10^{-15}$ & $( 8.312 \pm 0.468 \pm 2.201 ) \times 10^{-16}$ & \\
$6.705-6.7975$   & $( 2.244 \pm 0.068 \pm 0.409 ) \times 10^{-15}$ & $( 3.797 \pm 0.287 \pm 1.212 ) \times 10^{-16}$ & \\
$6.7975-6.89$    & $( 1.028 \pm 0.039 \pm 0.207 ) \times 10^{-15}$ & $( 3.045 \pm 0.192 \pm 0.869 ) \times 10^{-16}$ & \\
$6.89-6.9825$    & $( 5.513 \pm 0.241 \pm 1.232 ) \times 10^{-16}$ & $( 2.124 \pm 0.131 \pm 0.641 ) \times 10^{-16}$ & \\
$6.9825-7.075$   & $( 3.097 \pm 0.142 \pm 0.695 ) \times 10^{-16}$ & $( 1.418 \pm 0.086 \pm 0.421 ) \times 10^{-16}$ & \\
$7.075-7.1675$   & $( 1.560 \pm 0.053 \pm 0.346 ) \times 10^{-16}$ & $( 1.015 \pm 0.042 \pm 0.285 ) \times 10^{-16}$ & \\
$7.1675-7.26$    & $( 1.037 \pm 0.040 \pm 0.240 ) \times 10^{-16}$ & $( 6.321 \pm 0.299 \pm 1.791 ) \times 10^{-17}$ & \\
$7.26-7.3525$    & $( 6.769 \pm 0.252 \pm 1.533 ) \times 10^{-17}$ & $( 4.984 \pm 0.225 \pm 1.350 ) \times 10^{-17}$ & \\
$7.3525-7.445$   & $( 4.381 \pm 0.283 \pm 1.124 ) \times 10^{-17}$ & $( 3.040 \pm 0.196 \pm 0.897 ) \times 10^{-17}$ & \\
$7.445-7.5375$   & $( 2.277 \pm 0.126 \pm 0.549 ) \times 10^{-17}$ & $( 1.777 \pm 0.109 \pm 0.514 ) \times 10^{-17}$ & \\
$7.5375-7.63$    & $( 1.077 \pm 0.069 \pm 0.291 ) \times 10^{-17}$ & $( 8.163 \pm 0.585 \pm 2.629 ) \times 10^{-18}$ & \\
$7.63-7.7225$    & $( 7.968 \pm 0.754 \pm 2.301 ) \times 10^{-18}$ & $( 4.974 \pm 0.514 \pm 1.675 ) \times 10^{-18}$ & \\
$7.7225-7.815$   & $( 5.482 \pm 0.570 \pm 1.464 ) \times 10^{-18}$ & $( 3.423 \pm 0.398 \pm 0.986 ) \times 10^{-18}$ & \\
$7.815-7.9075$   & $( 2.426 \pm 0.249 \pm 0.633 ) \times 10^{-18}$ & $( 1.858 \pm 0.219 \pm 0.540 ) \times 10^{-18}$ & \\
$7.9075-8.0$     & $( 1.300 \pm 0.149 \pm 0.322 ) \times 10^{-18}$ & $( 1.104 \pm 0.143 \pm 0.300 ) \times 10^{-18}$ & \\
\hline
\end{tabular}
\end{table}

\begin{table}[htbp!]
\centering
\caption{
All-particle differential flux~($dJ/dE$) for our analysis, in $\left[ \text{m}^{-2} \text{sr}^{-1} \text{s}^{-1} \text{GeV}^{-1} \right]$.
``$\pm$\,Stat'' means statistical uncertainty of the flux, 
``$\pm$\,Bas.sys'' --- ``basic'' systematic uncertainties based on \mbox{QGSJet-II.04} model,
``Th. unc'' --- lower and upper theoretical uncertainties that cover results based on \mbox{QGSJet-II.04}, \mbox{EPOS-LHC} and \mbox{Sibyll 2.3c}.
}
\label{tab:total flux}
\begin{tabular}{|l|cccccc|}
\hline
 \makecell{ Energy bin, \\ $\log_{10}{(E/\text{GeV})}$ } & 
 \makecell{$dJ/dE$}  & $\pm$\,Stat & $\pm$\,Bas.\,sys &  \multicolumn{2}{c}{$-$ Th.\,unc $+$} &  \\
\hline
$6.15-6.2425$ & 5.738 & 0.012 & 0.931 & 1.384 & 1.073 & $\times 10^{-13}$ \\
$6.2425-6.335$ & 3.594 & 0.009 & 0.549 & 0.734 & 0.869 & $\times 10^{-13}$ \\
$6.335-6.4275$ & 2.189 & 0.006 & 0.342 & 0.555 & 0.404 & $\times 10^{-13}$ \\
$6.4275-6.52$ & 1.188 & 0.004 & 0.185 & 0.369 & 0.185 & $\times 10^{-13}$ \\
$6.52-6.6125$ & 6.073 & 0.029 & 0.929 & 1.476 & 1.287 & $\times 10^{-14}$ \\
$6.6125-6.705$ & 3.459 & 0.019 & 0.535 & 1.239 & 0.535 & $\times 10^{-14}$ \\
$6.705-6.7975$ & 1.855 & 0.013 & 0.286 & 0.550 & 0.307 & $\times 10^{-14}$ \\
$6.7975-6.89$ & 9.678 & 0.085 & 1.510 & 2.922 & 1.510 & $\times 10^{-15}$ \\
$6.89-6.9825$ & 5.267 & 0.057 & 0.834 & 1.512 & 0.834 & $\times 10^{-15}$ \\
$6.9825-7.075$ & 2.787 & 0.039 & 0.459 & 0.972 & 0.459 & $\times 10^{-15}$ \\
$7.075-7.1675$ & 1.238 & 0.024 & 0.209 & 0.397 & 0.209 & $\times 10^{-15}$ \\
$7.1675-7.26$ & 6.576 & 0.152 & 1.132 & 2.186 & 1.145 & $\times 10^{-16}$ \\
$7.26-7.3525$ & 3.929 & 0.101 & 0.684 & 1.275 & 1.175 & $\times 10^{-16}$ \\
$7.3525-7.445$ & 2.296 & 0.068 & 0.409 & 0.913 & 0.409 & $\times 10^{-16}$ \\
$7.445-7.5375$ & 1.096 & 0.041 & 0.201 & 0.491 & 0.230 & $\times 10^{-16}$ \\
$7.5375-7.63$ & 4.602 & 0.265 & 0.978 & 0.978 & 1.300 & $\times 10^{-17}$ \\
$7.63-7.7225$ & 2.818 & 0.199 & 0.628 & 1.016 & 1.337 & $\times 10^{-17}$ \\
$7.7225-7.815$ & 1.762 & 0.129 & 0.371 & 0.960 & 0.371 & $\times 10^{-17}$ \\
$7.815-7.9075$ & 9.231 & 0.891 & 1.991 & 2.557 & 1.991 & $\times 10^{-18}$ \\
$7.9075-8.0$ & 4.788 & 0.475 & 1.001 & 3.306 & 1.001 & $\times 10^{-18}$ \\
\hline
\end{tabular}
\end{table}

\clearpage

\bibliographystyle{JHEP}
\bibliography{biblio.bib}

\end{document}